\documentclass[9pt,conference]{IEEEtran}
\usepackage{amssymb,amsthm,amsmath,array}
\usepackage{graphicx}
\usepackage[caption=false,font=footnotesize]{subfig}
\usepackage{xspace}
\usepackage[sort&compress, numbers]{natbib}
\usepackage{stmaryrd}
\usepackage{xcolor}
\usepackage{mathtools}
\usepackage{float}
\usepackage{textcomp}

\newcommand{\ie}{\emph{i.e.}, }

\newcommand{\eint}[2]{E_1\big(\frac{#1}{#2}\big)}
\newcommand{\bs}{\boldsymbol}
\newcommand{\cl}{\mathcal}
\newcommand{\bb}{\mathbb}
\newcommand{\ud}{\mathrm{d}}
\usepackage{booktabs}  
\usepackage[mode=buildnew]{standalone}
\usepackage{tikz}

\usepackage{pgfplots}                     
\pgfplotsset{compat=1.18}                

\begin{document}
\title{Fast Simulation of Damage Diffusion Distribution in Scanning Transmission Electron Microscopy}
\author{\IEEEauthorblockN{
        Amir Javadi Rad\IEEEauthorrefmark{1},  
        Amirafshar Moshtaghpour\IEEEauthorrefmark{2},
        Dongdong Chen\IEEEauthorrefmark{1}, and
        Angus I. Kirkland\IEEEauthorrefmark{2}\IEEEauthorrefmark{3}
    }
    \IEEEauthorblockA{
        \IEEEauthorrefmark{1} Heriot-Watt University, Edinburgh, UK.\\
        \IEEEauthorrefmark{2} Rosalind Franklin Institute, Harwell Science \& Innovation Campus, Didcot, UK.\\
        \IEEEauthorrefmark{3} University of Oxford, Oxford, UK.}
}
\maketitle
\begin{abstract}
Scanning Transmission Electron Microscopy (STEM) is a critical tool for imaging the properties of materials and biological specimens at atomic scale, yet our understanding of relevant
electron beam damage mechanisms is incomplete. Recent studies suggest that certain types of damage can be modelled as a diffusion process.   However, numerical simulation of such diffusion processes has remained computationally intensive. This work introduces a high-performance C++ framework for simulating damage diffusion process in STEM that combines efficient numerical computation, advanced visualisations, and multithreading to achieve efficient runtime while maintaining accuracy. 
\end{abstract}

\section{Introduction}\label{sec:introduction}
Scanning Transmission Electron Microscopy (STEM) is a powerful technique for observing the nanoscale structure of materials~\cite{nellist1995resolution, james1999practical,zhang2018atomic}. In STEM, a convergent electron probe scans across the sample surface, which can lead to beam-induced damage. Common damage mechanisms include knock-on damage~\cite{egerton2004radiation}, caused by electron-atom interactions, and radiolysis ~\cite{egerton2019radiation}, which results from the cleavage of chemical bonds within the sample's structure.

While our understanding of the physical processes underlying these damage mechanisms remains limited, recent studies suggest that they exhibit spatio-temporal diffusion-like behaviour~\cite{jannis2022reducingpart2,nicholls2020minimising}.

Common methods to reduce beam damage in STEM rely on controlling the probe trajectory or subsampling probe positions. Alternating (or Interleaved) scans~\cite{interleaveSTEM2022} -- which skip fixed points on the grid -- and random scans~\cite{randomSTEM2020} -- where probe locations are selected randomly -- have demonstrated reduced damage compared to traditional raster scans. Inspired by the theory of compressive sensing~\cite{candes2006robust,donoho2006compressed}, subsampling electron probe positions followed by inpainting the missing probe positions data is another technique for mitigating damage~\cite{nicholls2023potential, nicholls2023targeted}. Non-conventional scanning strategies can be achieved with scan generators that alter the electron probe trajectory or beam blankers that selectively irradiate locations by rapidly deflecting the electron beam. Additionally, multi-frame fast acquisitions~\cite{multiframeSTEM2018} distribute the electron fluence across frames, reducing damage compared to single-frame acquisitions.

Previous studies~\cite{moshtaghpour2025diffusion,jannis2022reducingpart2} have employed diffusion models for qualitative analyses of electron beam damage during image acquisition. This work is built on the findings of~\cite{moshtaghpour2025diffusion}, which introduced the first explicit mathematical formulation of damage diffusion distribution in STEM. Despite the closed-form formulation of diffusion distribution provided in~\cite{moshtaghpour2025diffusion}, numerical simulations for STEM scans exceeding $32 \times 32$ probe positions remain impractical due to the high computational complexity involved. In this work, we present a C++ implementation of the diffusion model reported in~\cite{moshtaghpour2025diffusion} by leveraging powerful libraries for optimised computations and multithreading. Our implementation integrates the Intel Math Kernel Library (MKL) for efficient complex computations, Visualisation Toolkit (VTK) for visualisation of diffusion patterns, and Threading Building Blocks (TBB) for optimised multithreading. These tools enable scalable and computationally efficient simulations, allowing for the analysis of larger probe grids.

\section{Damage Diffusion Model} \label{sec:damage-diffusion-model}
Similar to~\cite{moshtaghpour2025diffusion}, we adopt the following assumptions:
\begin{itemize}
\item At each probe position, initial diffusing species are deposited at the rate of $Q_0$. Without specifying the exact relationship between $Q_0$ and the electron beam current, we assume that $Q_0$ (in $\rm u/\rm s$, for an arbitrary unit ''$\rm u$``) is independent of the dwell time, diffusion coefficient, and probe size.
\item The sample is uniform, infinitesimally thin, and significantly larger than the probe, allowing diffusion in STEM to be modelled as a two-dimensional (2D) process in an infinite medium.
\item The diffusion coefficient of the sample ($D$ in $\rm m^2/\rm s$) is constant.
\item The electron probe can be approximated by a Gaussian-shaped probe.
\item The number of diffusing species deposited by the electron probe is invariant with respect to changes in the probe size.
\end{itemize}
The diffusion distribution of an electron probe with radius $r_{\rm s}$ activated at location $\bs r_i$ and time $t_i$ for a duration (or dwell time) of $\tau_i$ can be modelled at locations $\bs r \ne \bs r_i$ as
\begin{equation}\label{eq:stem_single_probe_beamon}
    \phi_{i}^{\rm on}(\bs r,t) =
        \frac{Q_0}{4\pi D}\Big(\eint{\|\bs r - \bs r_i\|_2^2}{2D_s + 4D(t-t_i)}-\eint{\|\bs r - \bs r_i\|_2^2}{2D_s}\Big),
\end{equation}
when the beam is on, \ie $t_i \le t\le t_i + \tau_i$, and as
\begin{align}\label{eq:stem_single_probe_beamoff}
    \phi_i^{\rm off}(\bs r,t) =& 
        \frac{Q_0}{4\pi D}\Big(\eint{\|\bs r - \bs r_i\|_2^2}{2D_s +4D(t-t_i)} \nonumber\\&- \eint{\|\bs r - \bs r_i\|_2^2}{2D_s+4D(t-(t_i+\tau_i))}\Big), 
\end{align}
when the  beam is off, \ie $t>t_i+\tau_i$. In Eqs.~\eqref{eq:stem_single_probe_beamon} and \eqref{eq:stem_single_probe_beamoff}, $D_{\rm s} = r_{\rm s}^2$ and $E_1(v) \coloneqq \int_{v}^{+\infty} \frac{1}{u} e^{-u}\, \ud u$ for $v \in \bb{R}/\{0\}$ and $E_1(0) = +\infty$ is the Exponential integral of order one. . Moreover, the diffusion distribution at the corresponding probe position is 
\begin{align}\label{eq:dd_activation_point}
    \phi_{i}^{\rm on}(\bs r_i,t) &= 
        \frac{Q_0}{4\pi D}\ln{\big(\frac{D_s+2D(t-t_i)}{D_s}\big)}, \quad t_i \le t\le t_i + \tau_i\\
        \phi_i^{\rm off}(\bs r_i,t) &=
        \frac{Q_0}{4\pi D} \ln{\big(\frac{D_s+2D(t-t_i)}{D_s+2D(t-t_i-\tau_i)}\big)}, \quad t_i + \tau_i < t.
\end{align}
In the equations above, $\phi_{i}^{\rm on}$ and $\phi_{i}^{\rm off}$ are in $\rm u/\rm m^2$. Moreover, these functions are the solution of the second Fick's model \cite[\textsection 1.2]{crank1979mathematics}:
\begin{equation}\label{eq:general-pde-constant-d}
    \frac{\partial \phi_i(\bs r, t)}{\partial t} = D\big(\frac{\partial^2 \phi_i(\bs r, t)}{\partial r_1^2} + \frac{\partial^2 \phi_i(\bs r, t)}{\partial r_2^2} \big).
\end{equation}

\begin{figure}
    \centering
    \begin{minipage}{\textwidth}
    \begin{minipage}{0.245\textwidth}
        \includegraphics[width=1\linewidth, trim = {2.2cm 1cm 1.5cm 1cm}, clip]{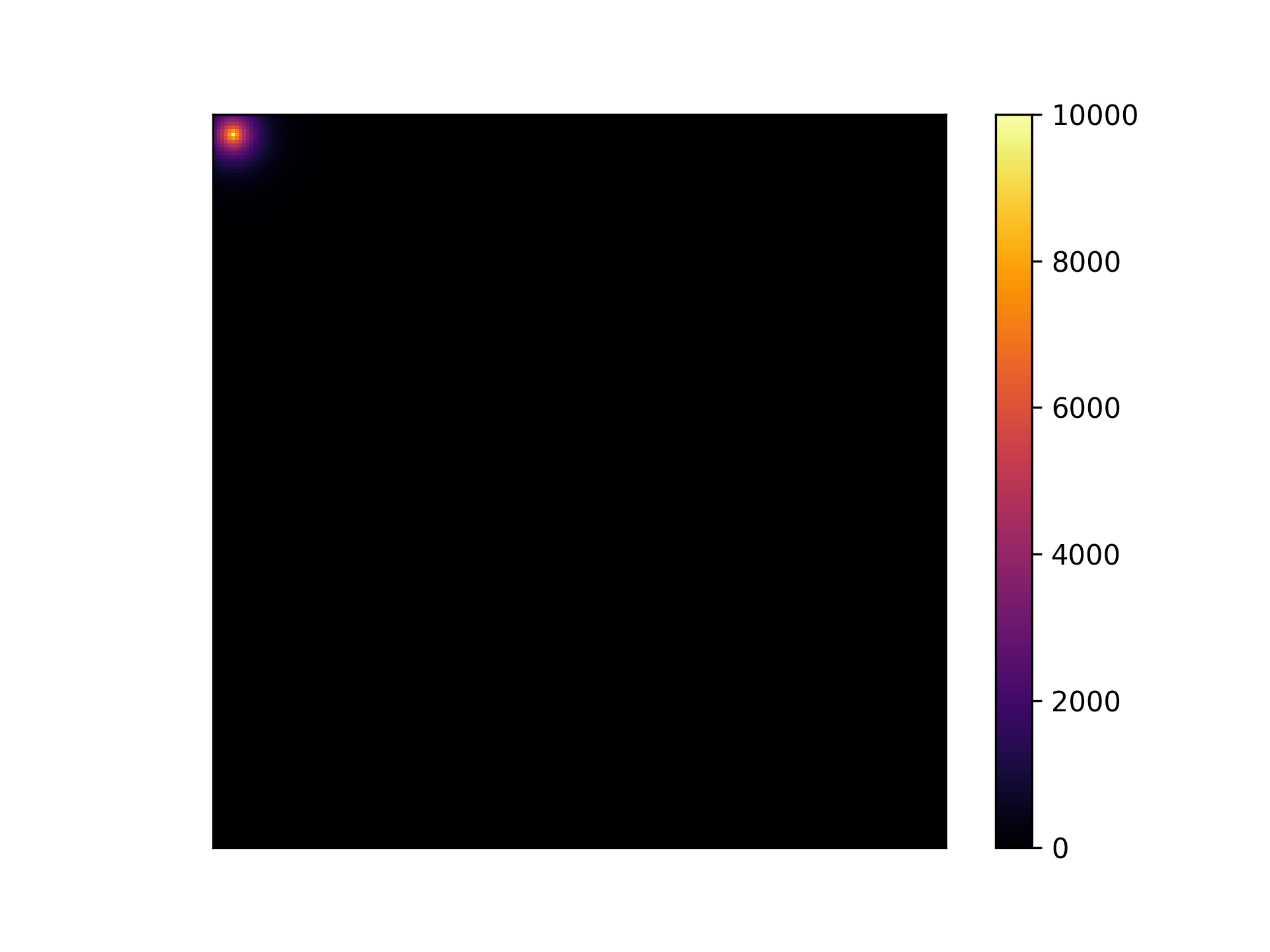}
    \end{minipage}
    \begin{minipage}{0.245\textwidth}
        \includegraphics[width=1\linewidth, trim = {2.2cm 1cm 1.5cm 1cm}, clip]{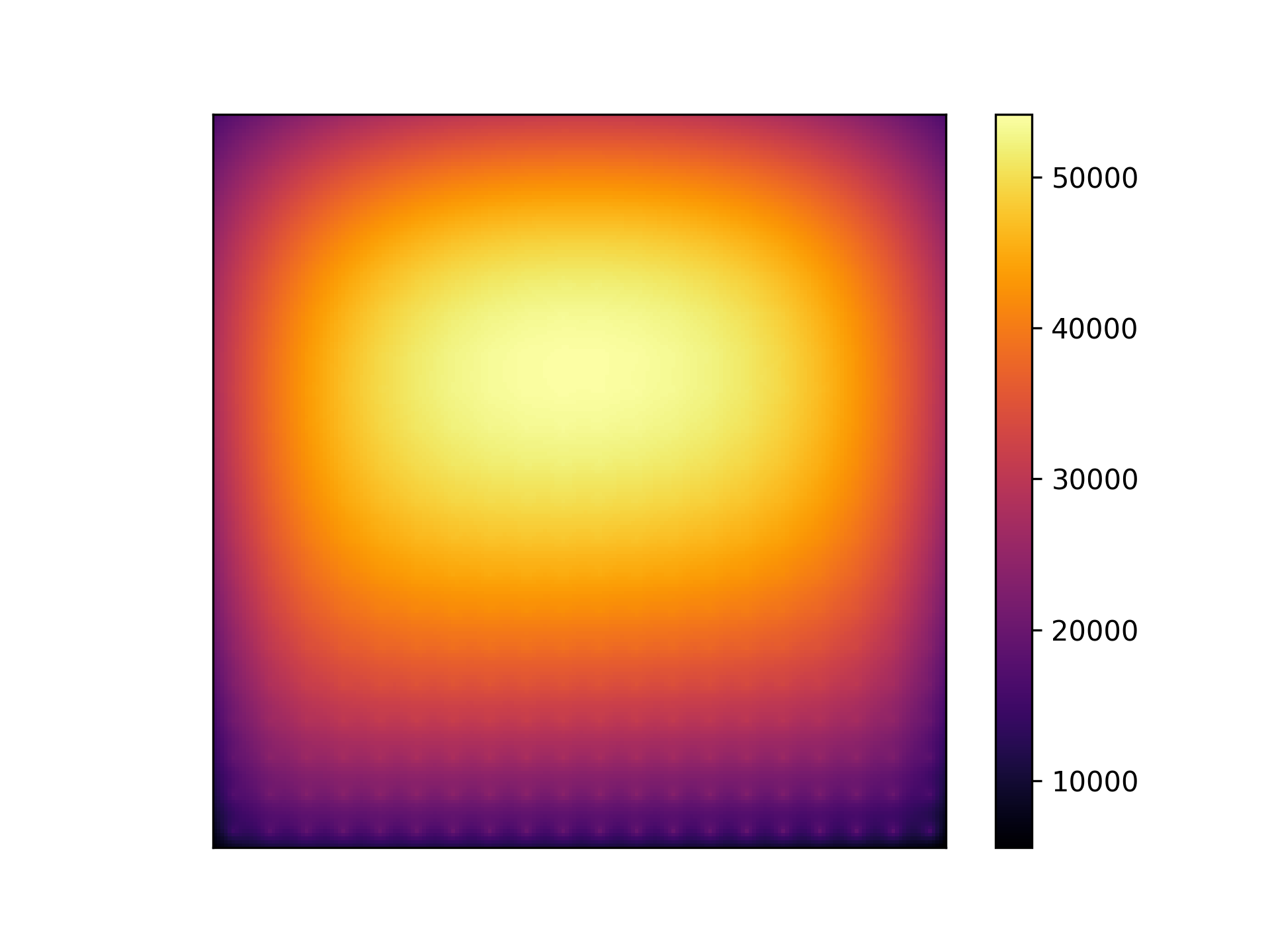}
    \end{minipage}
    \end{minipage}
    \caption{Cumulative diffusion distribution examples at the beginning (left) and end (right) of the scan.}
    \label{fig:examples}
\end{figure}
\begin{figure}[tb]
    \centering
\includegraphics[width=.4\textwidth]{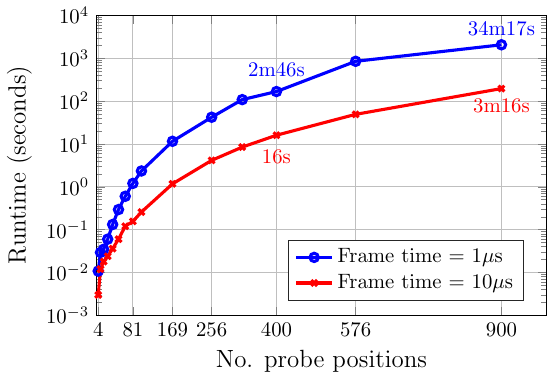}
    \caption{Simulation execution time using our C++ framework for various number of probe positions for two frame time values.}
    \label{fig:runtime_curve}
\end{figure}

In STEM, probe positions are sequentially scanned, one after the other, requiring us to account for the cumulative effect of the diffusion distribution. Accordingly, the cumulative diffusion distribution resulting from scanning $j$ probe positions is expressed as:
\begin{equation}\label{eq:stem_cumulative_simple}
    \psi_j(\bs r, t) \coloneqq \sum_{i=1}^{j} \phi_i(\bs r, t).
\end{equation}
\section{Computational Complexity} \label{sec:computational-complexity}
As reported in~\cite{moshtaghpour2025diffusion}, the time complexity of computing the cumulative diffusion distribution, as formulated in \eqref{eq:stem_cumulative_simple}, for a full STEM scan with $N$ probe positions is
\begin{equation}\label{eq:computational-omplexity}
   \cl O(\delta^{-1}_{\rm t} \delta^{-2}_{\rm s} N^3 T_{E_1}),
\end{equation}
where $\delta_{\rm t}$ is the frame time, $\delta_{\rm s}$ is the pixel side length of the simulation, and $T_{E_1}$ is a time complexity of $E_1$ function. It is evident from Eq.~\eqref{eq:computational-omplexity} that the time complexity of the cumulative diffusion distribution scales with $N^3$. This highlights that the primary bottleneck in simulating damage diffusion distribution in STEM lies in the number of probe positions.
To address this computational challenge, we implemented the diffusion model in C++, which inherently offers lower-level memory management and efficient resource handling, resulting in faster execution compared to higher-level interpreted languages like Python. These optimisations are critical for large-scale computations where reducing runtime complexity is essential. Our work also focused on designing an efficient multithreaded approach tailored to this problem's specific requirements. 
We partitioned the computation into independent tasks, each processing subsets of the probe positions concurrently. Special care was taken to balance the computational load across threads, ensuring that no single thread was overloaded. 


\begin{figure}[htbp]
  \centering
  \includegraphics[width=0.35\textwidth]{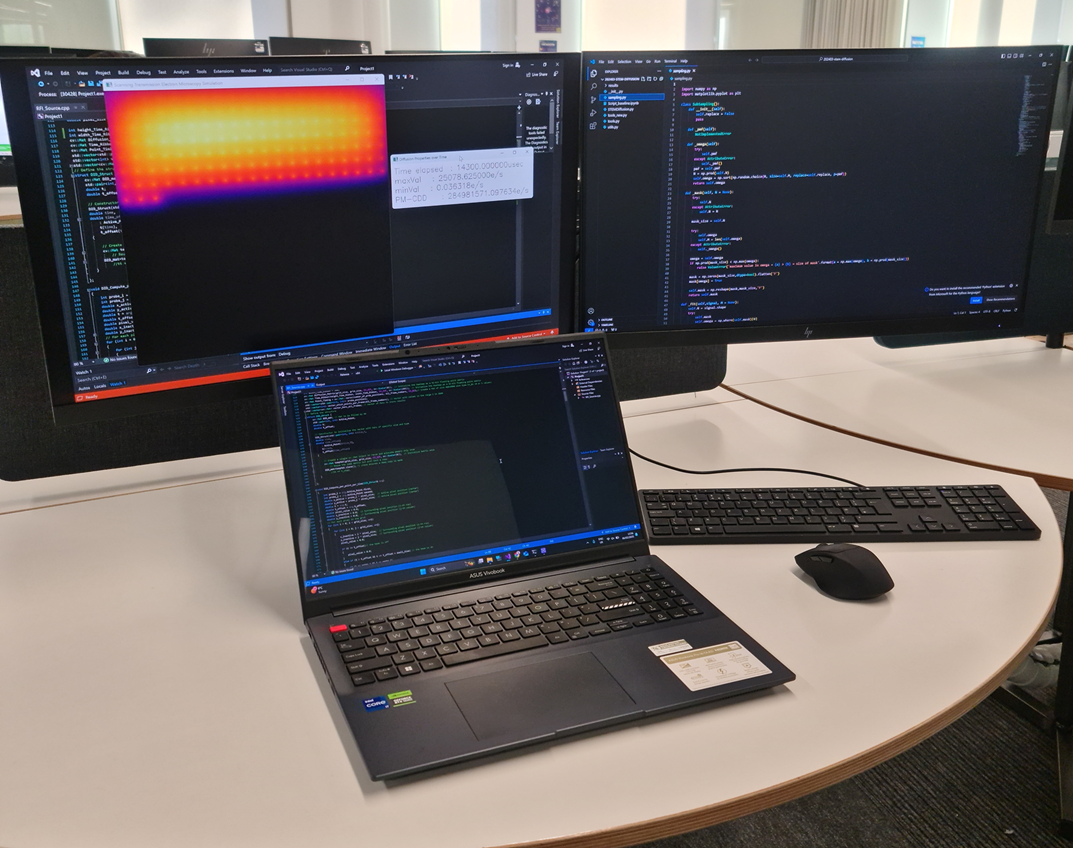}
  \caption{Our setup for show-and-tell demo.}
  \label{fig:setup}
\end{figure}

\section{Library Features} \label{sec:library-features}
Our framework leverages the following libraries:
\begin{itemize}
    \item the Intel Math Kernel Library (MKL) for highly optimised mathematical operations, enabling precise and efficient computation of complex diffusion equations;
    \item the Visualization Toolkit (VTK) for integrated live or offline visualisation, providing comprehensive 2D and 3D graphical representations of damage diffusion process;
    \item the Intel Threading Building Blocks (TBB) for multithreading, allowing for a robust library for scalable parallel programming with efficient utilisation of multi-core processors and significant runtime reductions.
\end{itemize}

By leveraging C++'s low-level control and incorporating powerful libraries, the proposed framework minimizes computational overhead, accelerates large-scale simulations, and facilitates detailed analyses of diffusion phenomena. This approach significantly improves runtime performance compared to Python implementations, establishing it as a valuable tool for advancing research in electron microscopy for materials science and structural biology. For instance, while a Python implementation of the cumulative diffusion distribution for a full STEM scan with $20 \times 20$ probe positions and a $10\mu{\rm s}$ frame time takes approximately 50 minutes, our C++ framework achieves the same task in just 16 seconds, as illustrated in Fig.~\ref{fig:runtime_curve}.

\section{Visitors Experience} \label{sec:visitors-experience}
Attendees will have the opportunity to modify simulation parameters, providing hands-on experience with the library. 
The demonstration will be conducted using a desktop computer -- see our setup in Fig.~\ref{fig:setup}, a monitor, and a poster that includes supporting information on the working principles of STEM and the diffusion model.

\section{Conclusion}
We plan to showcase a C++ framework for fast simulation of damage diffusion distribution in STEM. Our implementation enables simulating the cumulative diffusion distribution of a full STEM scan with $20 \times 20$ probe positions in 16 seconds, significantly faster than the Python implementation, which took approximately 50 minutes. This performance was achieved by utilising powerful libraries, including MKL (for efficient complex computations), VTK (for integrated visualisation), and TBB (for multithreading). In the future, we plan to extend this framework into a GPU-parallelised C++ library to enable real-time simulations of damage diffusion distribution.
\bibliographystyle{IEEEtran}
\bibliography{references}

\begin{thebibliography}{10}
\providecommand{\url}[1]{#1}
\csname url@samestyle\endcsname
\providecommand{\newblock}{\relax}
\providecommand{\bibinfo}[2]{#2}
\providecommand{\BIBentrySTDinterwordspacing}{\spaceskip=0pt\relax}
\providecommand{\BIBentryALTinterwordstretchfactor}{4}
\providecommand{\BIBentryALTinterwordspacing}{\spaceskip=\fontdimen2\font plus
\BIBentryALTinterwordstretchfactor\fontdimen3\font minus \fontdimen4\font\relax}
\providecommand{\BIBforeignlanguage}[2]{{%
\expandafter\ifx\csname l@#1\endcsname\relax
\typeout{** WARNING: IEEEtran.bst: No hyphenation pattern has been}%
\typeout{** loaded for the language `#1'. Using the pattern for}%
\typeout{** the default language instead.}%
\else
\language=\csname l@#1\endcsname
\fi
#2}}
\providecommand{\BIBdecl}{\relax}
\BIBdecl

\bibitem{nellist1995resolution}
P.~Nellist, B.~McCallum, and J.~M. Rodenburg, ``Resolution beyond the information limit in transmission electron microscopy,'' \emph{nature}, vol. 374, no. 6523, pp. 630--632, 1995.

\bibitem{james1999practical}
E.~James and N.~Browning, ``Practical aspects of atomic resolution imaging and analysis in {STEM},'' \emph{Ultramicroscopy}, vol.~78, no. 1-4, pp. 125--139, 1999.

\bibitem{zhang2018atomic}
D.~Zhang, Y.~Zhu, L.~Liu, X.~Ying, C.-E. Hsiung, R.~Sougrat, K.~Li, and Y.~Han, ``Atomic-resolution transmission electron microscopy of electron beam--sensitive crystalline materials,'' \emph{Science}, vol. 359, no. 6376, pp. 675--679, 2018.

\bibitem{egerton2004radiation}
R.~Egerton, P.~Li, and M.~Malac, ``Radiation damage in the {TEM} and {SEM},'' \emph{Micron}, vol.~35, no.~6, pp. 399--409, 2004.

\bibitem{egerton2019radiation}
R.~Egerton, ``Radiation damage to organic and inorganic specimens in the {TEM},'' \emph{Micron}, vol. 119, pp. 72--87, 2019.

\bibitem{jannis2022reducingpart2}
D.~Jannis, A.~Velazco, A.~Béché, and J.~Verbeeck, ``Reducing electron beam damage through alternative {STEM} scanning strategies, part ii: Attempt towards an empirical model describing the damage process,'' \emph{Ultramicroscopy}, vol. 240, p. 113568, 2022.

\bibitem{nicholls2020minimising}
D.~Nicholls, J.~Lee, H.~Amari, A.~J. Stevens, B.~L. Mehdi, and N.~D. Browning, ``Minimising damage in high resolution scanning transmission electron microscope images of nanoscale structures and processes,'' \emph{Nanoscale}, vol.~12, no.~41, pp. 21\,248--21\,254, 2020.

\bibitem{interleaveSTEM2022}
A.~Velazco, A.~Béché, D.~Jannis, and J.~Verbeeck, ``Reducing electron beam damage through alternative {STEM} scanning strategies, part i: Experimental findings,'' \emph{Ultramicroscopy}, vol. 232, p. 113398, 2022.

\bibitem{randomSTEM2020}
A.~Zobelli, S.~Y. Woo, A.~Tararan, L.~H.~G. Tizei, N.~Brun, X.~Li, O.~Stéphan, M.~Kociak, and M.~Tencé, ``Spatial and spectral dynamics in {STEM} hyperspectral imaging using random scan patterns,'' \emph{Ultramicroscopy}, vol. 212, p. 112912, 2020.

\bibitem{candes2006robust}
E.~Cand\'es, J.~Romberg, and T.~Tao, ``Robust uncertainty principles: Exact signal reconstruction from highly incomplete frequency information,'' \emph{IEEE Transactions on information theory}, vol.~52, no.~2, pp. 489--509, 2006.

\bibitem{donoho2006compressed}
D.~L. Donoho, ``Compressed sensing,'' \emph{IEEE Transactions on Information Theory}, vol.~52, pp. 1289--1306, 2006.

\bibitem{nicholls2023potential}
D.~Nicholls, M.~Kobylysnka, J.~Wells, Z.~Broad, D.~McGrouther, A.~Moshtaghpour, , A.~I. Kirkland, R.~A. Fleck, and N.~D. Browning, ``The potential of subsampling and inpainting for fast low-dose cryo {FIB-SEM} imaging and tomography,'' \emph{arXiv preprint arXiv:2309.09617}, 2023.

\bibitem{nicholls2023targeted}
D.~Nicholls, J.~Wells, A.~W. Robinson, A.~Moshtaghpour, M.~Kobylynska, R.~A. Fleck, A.~I. Kirkland, and N.~D. Browning, ``A targeted sampling strategy for compressive cryo focused ion beam scanning electron microscopy,'' in \emph{ICASSP 2023-2023 IEEE International Conference on Acoustics, Speech and Signal Processing (ICASSP)}.\hskip 1em plus 0.5em minus 0.4em\relax IEEE, 2023, pp. 1--5.

\bibitem{multiframeSTEM2018}
L.~Jones, A.~Varambhia, R.~Beanland, D.~Kepaptsoglou, I.~Griffiths, A.~Ishizuka, F.~Azough, R.~Freer, K.~Ishizuka, D.~Cherns, Q.~M. Ramasse, S.~Lozano-Perez, and P.~D. Nellist, ``Managing dose-, damage- and data-rates in multi-frame spectrum-imaging,'' \emph{Microscopy}, vol.~67, p. i98–i113, 2018.

\bibitem{moshtaghpour2025diffusion}
A.~Moshtaghpour, A.~Velazco-Torrejon, D.~Nicholls, A.~W. Robinson, A.~I. Kirkland, and N.~D. Browning, ``Diffusion distribution model for damage mitigation in scanning transmission electron microscopy,'' \emph{Journal of Microscopy}, vol. 297, no.~1, pp. 57--77, 2025.

\bibitem{crank1979mathematics}
J.~Crank, \emph{The mathematics of diffusion}.\hskip 1em plus 0.5em minus 0.4em\relax Oxford university press, 1979.

\end{thebibliography}
\end{document}